\documentclass[12pt,a4paper]{article}
\usepackage{graphicx}
\usepackage{geometry}
\usepackage{amsmath}
\usepackage{amssymb}
\usepackage{epstopdf}
\usepackage{hyperref}
\usepackage{chngpage}
\usepackage{subfigure}
\usepackage{float}
\usepackage{color,pictexwd,psfrag}
\restylefloat{table}

\newcommand{\figs}{draft_fig/}

\definecolor{blue}{rgb}{0.0,0.0,0.0} % blue for reviewer 1 
\definecolor{green}{rgb}{0.0,0.0,0.0} % green for reviewer 3
\definecolor{red}{rgb}{0.0,0.0,0.0} % red for reviewer 4 
\definecolor{purple}{rgb}{0.0,0.0,0.0}
\definecolor{orange}{rgb}{0.0,0.0,0.0}

\begin{document}

\begin{center}
\LARGE{\textbf{Assessment of deconvolution-based flamelet methods for progress variable rate modeling.}}
\end{center}

\vspace{0.5cm}

\textbf{Z.M. Nikolaou$^1$, L. Vervisch$^2$}

\vspace{0.5cm}

$^1$Computation-based Science and Technology Research    Centre (CaSToRC), The Cyprus Institute, Nicosia, 2121, Cyprus. z.nicolaou@cyi.ac.cy
           
$^2$CORIA-CNRS, Normandie Universite, INSA de Rouen  Normandie, Saint-Etienne-du-Rouvray, France.   vervisch@coria.fr
      
\vspace{0.5cm}
      
\textbf{Correspondence:} z.nicolaou@cyi.ac.cy
              
\vspace{0.5cm}

\noindent \textbf{Abstract}

\vspace{0.5cm}

\noindent A novel approach for modeling the progress variable reaction rate in Large Eddy Simulations of turbulent and reacting flows is proposed. This is done in the context of two popular flamelet models which require the progress variable variance as input. The approach is based on using a recently proposed deconvolution method for modeling the variance. The deconvolution-modeled variance, is used as an input in the flamelet models for modeling the filtered progress variable rate. The assessment of the proposed approach is conducted a priori using direct numerical simulation data of turbulent premixed flames. For the conditions tested in this study, deconvolution does not introduce a significant bias in the flamelet models' predictions, while a quantitatively good prediction of the progress variable rate is obtained for both flamelet models considered. 

%---------------------------------------------------------------------------

\section{Introduction}

Large Eddy Simulation (LES) is becoming the de facto approach both in industry and academia for modeling a wide range of flows. This includes a wide range of devices with complex geometry in the aerospace, energy, marine, and automotive industries to name but a few. The numerical modeling however of turbulent and reacting flows is computationally challenging and expensive, particularly in the context of LES. Reacting flows, and specifically combustion processes, include a wide range of length and time scales, all of which need to be adequately resolved during the simulation, on a mesh which for practical geometries can include millions of cells \cite{sagaut_book_2001}. An accurate description of the chemical processes involved in such flows, requires the use of fairly large detailed chemical kinetics mechanisms with hundreds of species and reactions which poses a significant computational workload. 

In an effort to reduce the dimensionality of the problem, various combustion modeling strategies have been developed throughout the years. The earliest modeling attempts employed simple, 1-step chemistry, for modeling the chemical kinetics \cite{bray_aast_1977}. This, reduces the computational load substantially. However, with an ever increasing need for accuracy but at the same time detailed chemistry information (e.g. pollutants concentrations, $\mathrm{CO_2}$, $\mathrm{NO_x}$ etc.), more elaborate combustion models were developed able to describe complex chemistry effects. These include, among others, Thickened Flames (TF), Flamelet Methods (FM), Conditional Moment Closure (CMC), transported-pdf methods etc. Each method has both its merits and drawbacks, and an in-depth discussion on the subject can be found in the reviews \cite{gicquel_pec_2012}-\cite{pitsch_annrev_2006}.
 
Flamelet methods in particular are popular since they are relatively straightforward to implement, and have been applied in a number of LES studies both for premixed and for non-premixed flames, with overall good results \cite{domingo_pof_2008}-\cite{langella_cnf_2016}. These studies include flows in complex geometries and for different flow configurations (freely-propagating, recirculating, etc.). The filtered progress variable $\tilde{c}$ which is a main LES solution variable, and it's variance ${\sigma}^2=\widetilde{c^2}-\tilde{c} \tilde{c}$ as obtained from a suitable model, are used as parameters in pre-built tables for obtaining important information and for providing closures for un-closed terms in the governing equations. This includes species mean (filtered) mass fractions, mean (filtered) reaction rates etc. Such methods offer a simple, yet effective way of providing detailed chemistry information (a posteriori), while keeping the computational requirements to a minimum. 

A key unclosed term in the transport equation for the progress variable, $\tilde{c}$, is the filtered progress variable reaction rate, $\bar{\dot{w}} _c$, where the overbar denotes a spatial filtering operation as defined in the context of LES. This is a dominant term, particularly in the reaction zone of the flame \cite{boger_proccomb_1998}, and an accurate model is required for this term. Two popular flamelet methods for modeling $\bar{\dot{w}} _c$, and which involve the progress variable variance, are the presumed-pdf approach and the Filtered Laminar Flame (FLF) approach \cite{moureau_cnf_2010,nambully_cnf_2014}. In the presumed-pdf approach, the progress variable $\tilde{c}$ and it's variance $\sigma ^2$ , are used to parameterise the progress variable pdf $\tilde{p}(\zeta;\tilde{c},\sigma ^2)$, where $\zeta$ is the sample-space variable for $\tilde{c}$. This pdf is usually taken to be a $\beta$-function. A usually steady, 1D, laminar flame solution is used for integrating with the pdf thus obtaining a closure for the rate. In the FLF approach, a 1D laminar flame solution is pre-filtered, and a table is constructed which contains values of the filtered reaction rate (or other variables of interest). This is done for discrete filter-width values, and the table contains the filtered values of the variables at all spatial positions across the 1D flame. The progress variable as obtained from the LES, and its variance, are used as parameters for accessing the table, and obtaining a value of the filtered rate. It is clear that in both methods the progress variable variance is an important parameter for determining the filtered rate. In turn, the value of the rate affects the flame propagation characteristics (flame displacement speed), and the interaction of the flame with the flow field \cite{boger_proccomb_1998}.

In the traditional modeling approach, an algebraic model is used or a transport equation is solved for in order to obtain an estimate of the variance, and there exist many studies in the literature using these two classic modeling approaches \cite{cook_pof_1994}-\cite{balarac_pof_2008}. Algebraic-based models are generally computationally efficient, whereas an additional transport equation requires solving in the case of transport equation-based models. Both of these approaches  however, usually include additional unknown model parameters whose value significantly affects the predictive ability of the model. In classic, gradient-based methods for example, a filter-size dependent constant is involved \cite{girimaji_pof_1996}. In transport-equation based methods, additional un-closed terms appear in the transport equation for the variance which also require modeling-an important un-closed term is the scalar dissipation rate. In addition, the value of the models' parameters depend on many factors such as the flow configuration, Reynolds number $Re$, Damkohler number $Da$ etc. This fact, may limit the generality of classic modeling approaches, and as a result the applicability range of flamelet methods which require the variance as input. 

An alternative modeling approach in the context of LES is based on deconvolution. Deconvolution-based methods aim to recover a good approximation $\phi ^*$ of the un-filtered field $\phi$ from knowledge of the filtered field $\bar{\phi}$ on the LES mesh. Deconvolution was recently found to be a powerful tool to complement physical modeling by directly exploiting the information contained in the  resolved signals \cite{bose_pof_2014,locci_ftac_2016}. In reacting flows, deconvolution methods were used in \cite{mathew_proc_2002,vreman_ftac_2009}, and more recently in \cite{domingo_cnf_2017}-\cite{wang_cnf_2017} in order to recover variables such as the temperature, with overall promising results in both a priori and a posteriori studies.  

In recent work \cite{nikolaou_ftc_2018,nikolaou_prf_2018}, an iterative deconvolution method was proposed which was shown to provide quantitatively good predictions of the un-filtered progress variable. Furthermore, it was shown that Iterative Deconvolution combined with Explicit Filtering (IDEF) provides quantitatively good predictions of the progress variable variance and the scalar flux \cite{nikolaou_ftc_2018,nikolaou_prf_2018}. Given that diffusive terms in the transport equation for the progress variable can be adequately modeled with standard practices, the only remaining term is the filtered progress variable reaction rate. A hybrid deconvolution-flamelet formulation is thus proposed for the above two flamelet approaches, for modeling the filtered rate. The main advantages of such an approach are twofold: (a) detailed chemistry information is provided a posteriori while still solving a single transport equation for the progress variable, and (b) a parameter-free method, namely deconvolution, which is not tied to a specific flow regime or combustion mode, is used for obtaining the variance. These two advantages combined, have the potential to render a more robust flamelet strategy for modeling the reaction rate in turbulent premixed flames in the context of LES. 

Since the deconvolution method described in \cite{nikolaou_ftc_2018,nikolaou_prf_2018} is local i.e. the method provides a value of the variance at every mesh point, it is still unclear what the overall effect of modeling the progress variable reaction rate will be when using a flamelet method such as a presumed-pdf approach, or FLF. The aim of this study, is to evaluate the sensitivity in modeling the filtered rate using these two approaches, where IDEF is employed for modeling the variance. In contrast to most a priori assessments which are conducted on the fine DNS mesh, the assessment in this study is conducted on a much coarser LES mesh in order to simulate an actual LES. Even though a priori assessments in general using DNS data do not guarantee functionality of the model in actual LES, the results in \cite{nikolaou_ftc_2018,nikolaou_prf_2018} indicate that the simulated LES a priori assessment procedure is more suitable for evaluating LES sub-grid models. This is particularly true for sub-grid scale models which include gradients of the resolved fields \cite{nikolaou_ftc_2018}. In addition, such a priori assessments are useful in clearly evaluating the performance of a combustion model, free from the influences of turbulence models and of the often dissipative numerical schemes which are used in actual LES. 

In the text which follows, section \ref{sec:DNS dat} lists details of the DNS database used for the assessment, section \ref{sec:mathback} describes the hybrid deconvolution-flamelet approach used, and results are presented/discussed in section \ref{sec:resdisc}.

\section{Description of the DNS database}\label{sec:DNS dat}

Direct simulations were conducted using the SENGA2 code \cite{senga2}. The code solves the compressible Navier-Stokes equations using a 10th order finite difference scheme for interior points, and a 4th order Runge-Kutta scheme for the time-stepping,

\begin{equation}
\frac{\partial\rho}{\partial t} +
\frac{\partial\rho u_{k}}{\partial x_{k}} = 0\:,
\label{CONTINUITY}
\end{equation}

%the Navier-Stokes momentum equation:
\begin{equation}
\frac{\partial\rho u_{i}}{\partial t} +
\frac{\partial\rho u_{k}u_{i}}{\partial x_{k}} =
-\frac{\partial p }{\partial x_{i}}+
\frac{\partial \tau_{ki} }{\partial x_{k}}\:,
\end{equation}

%the internal energy equation:
\begin{equation}
\frac{\partial\rho E}{\partial t} +
\frac{\partial\rho u_{k} E }{\partial x_{k}} =
-\frac{\partial p u_{k}}{\partial x_{k}} -
\frac{\partial q_{k}}{\partial x_{k}}  +
\frac{\partial \tau_{km}u_{m}}{\partial x_{k}} \:,
\end{equation}

%and the species mass fraction equations:

\begin{equation}
\frac{\partial\rho Y_{\alpha}}{\partial t} +
\frac{\partial\rho u_{k} Y_{\alpha}}{\partial x_{k}} =
\dot w_{\alpha} -
\frac{\partial\rho V_{\alpha,k} Y_{\alpha} }{\partial x_{k}}\:.
\label{MASSFRAC}
\end{equation}
 
\noindent where the symbols have their usual meanings. Table \ref{tbl:turb_param} lists details of the DNS database used to conduct the analysis. This is a freely-propagating premixed multi-component fuel-air flame. A detailed chemical mechanism which was developed specifically for such fuels was used with 49 reactions and 15 species \cite{nikolaou_cnf_2013}. $u_{rms}$ is the rms value of the fluctuating component of the incoming velocity field, with an integral length scale $l_{int}$ on the reactant side. The turbulence Reynolds number is $Re=u_{rms}l_{int}/ \nu _r$, the Damkohler number is $Da=(l_{int}/u_{rms})/(\delta / s_L)$ and the Karlovitz number is $Ka=(\delta/ \eta _k)^2$, where $s_L$ is the laminar flame speed, and the thickness $\delta=\nu _r/s_L$. Note that the laminar flame thickness is defined as $\delta _L$=$(T_p-T_r)/max(dT/dx)$ where $T_r$, $T_p$ are the reactant and product temperatures respectively. These conditions place the flame in the distributed/broken reaction zones regime according to the classic combustion diagram by Peters \cite{peters_symp_1986}. Further details of the simulations are given in \cite{nikolaou_cnf_2014,nikolaou_cst_2015}. In the sections which follow, the progress variable $c$ is based on temperature.

\begin{table}[h!]
\centering
\begin{tabular}{ l c c c c c c c c c}
\hline
 Case & $u_{rms}/s_l$ & $l_{int}/{\delta}$ & $Ret$ & $Da$ & $Ka$ \\ 
\hline
 A &3.18  &16.54 &52.66  &5.19 &1.39   \\
 B &9.00  &16.66 &150.05 &1.85 &6.62   \\
 C &14.04 &16.43 &230.69 &1.17 &12.97  \\
\hline
\end{tabular}
\caption{Turbulent flame parameters for the DNS studies.}
\label{tbl:turb_param}
\end{table}

\section{Mathematical background}\label{sec:mathback}

\subsection{Filtering and deconvolution operations}

In the context of LES, filtering is a convolution operation between the signal $\phi$ and the filter $G$, 

\begin{equation}\label{eq:filtop}
\bar{\phi}(\underline{x},t)=\int_{\underline{x}'=-\infty} ^{\infty}G(\underline{x}-\underline{x}') \phi(\underline{x}',t)d\underline{x}'=G*\phi
\end{equation}

\noindent A Gaussian filter is used in the present study, 

\begin{equation}
G(\underline{x})=\left( {\frac{6}{\pi {\Delta} ^{2}}} \right)^{\frac{3}{2}}  e^{\frac{-6\underline{x} \cdot \underline{x}}{{\Delta} ^{2}}}
\end{equation}

\noindent where $\Delta$ is the filter width. Favre-filtered variables are defined as,

\begin{equation}
\tilde{\phi}(\underline{x},t)=\frac{\overline{\rho \phi}}{\bar{\rho}}
\end{equation} 

\noindent The DNS data $\rho$ and $\rho c$ are first filtered using Eq. \ref{eq:filtop} on the fine DNS mesh. This is done for filter widths $\Delta ^+=\Delta / \delta _L$ of 1.0, 2.0 and 3.0 respectively. In order to simulate an actual LES, the filtered data as obtained on the DNS mesh are then sampled onto a much coarser LES mesh. This is done using high-order Lagrange polynomials for the interpolation. The filtered variables as obtained by sampling on the LES mesh are then $\bar{\rho}$ and $\overline{\rho c}$. The size of the LES mesh is determined based on the criterion described in \cite{nikolaou_ftc_2018}: the deconvolution method requires the filtered fields to be adequately resolved on the LES mesh, and the criterion derived in \cite{nikolaou_ftc_2018} ensures this. This requirement is translated to an LES mesh spacing ratio $h/ \Delta$=0.25 where $h$ is the LES mesh spacing. Tables \ref{tbl:mesh_A} and \ref{tbl:mesh_B} list the DNS and LES meshes for each of the three DNS databases used in this study. In comparison to the Kolmogorov length scale $\eta _k$, the LES mesh is much coarser: $h/\eta _k$ for cases A, B and C for the finest LES mesh ($\Delta ^+=$1.0) is 6.8, 14.9 and 20.7 respectively, and much larger for the coarsest LES mesh considered in this study. Therefore, small-scale information of the order of $\eta _k$ is lost on the simulated LES mesh. 

\begin{table}[h!]
\centering
\begin{tabular}{l c c c c}
\hline
$\Delta ^+$ & $N_x$ & $N_y$ & $N_z$ & $h$/ $l_T$  \\ [0.75ex]
\hline
DNS & 768 &384 &384 & 0.03 \\
1 & 74  &37  &37  & 0.35\\  
2 & 37  &18  &18  & 0.69\\
3 & 24  &12  &12  & 1.04\\
\hline
\hline
\end{tabular}
\caption{DNS and LES meshes for cases A  and B with $h / \Delta=0.25$.}
\label{tbl:mesh_A}
\end{table}

\begin{table}[h!]
\centering
\begin{tabular}{l c c c c}
\hline
$\Delta ^+$ & $N_x$ & $N_y$ & $N_z$ & $h$/ $l_T$ \\ [0.75ex]
\hline
DNS & 1632 &544 &544 & 0.02\\
1 & 112  &37  &37  & 0.21\\ 
2 & 56  &18  &18   & 0.42\\
3 & 37  &12  &12   & 0.64\\
\hline
\hline
\end{tabular}
\caption{DNS and LES meshes for case C with $h / \Delta=0.25$.}
\label{tbl:mesh_B}
\end{table}

The filtered variables $\bar{\rho}$  and $\overline{\rho c}$ as obtained by sampling on the LES mesh are deconvoluted using an iterative algorithm in order to obtain estimates of the un-filtered fields namely $\rho ^*$ and $( \rho c ) ^*$ (on the LES mesh). The algorithm (Van-Cittert) reads \cite{jansson_ny_1984}-\cite{benjamin_ieee_1991},

\begin{equation}\label{eq:deconvVC}
{\phi ^*}^{n+1}={\phi ^*}^{n}+b(\bar{\phi}-G*{{\phi ^*}^n})\:.
\end{equation} 

\noindent with $\phi ^{*0}=\bar{\phi}$, and the progress variable variance is calculated by explicitly filtering the deconvoluted fields as obtained on the LES mesh, 

\begin{equation}\label{eq:varmdl}
{\sigma ^2}_{IDEF}=\frac{\overline{(\rho c)^* (\rho c)^* / \rho ^*}}{\bar{\rho}}-\tilde{c} \tilde{c}
\end{equation}

\noindent In practice, the deconvolution algorithm is implemented with an error controller in order to ensure that the deconvolved fields lie within correct physical limits. Further details of the exact implementation of the method are given in \cite{nikolaou_ftc_2018,nikolaou_prf_2018}. The estimate of the variance as given by Eq. \ref{eq:varmdl} is then used for parameterising the reaction rate models which are presented in the next two sections. 

\subsection{Unstrained Flamelet presumed-pdf (UF) model}{\label{sec:uf}}

In the classical UF modeling approach the filtered rate is calculated using, 

\begin{equation}\label{eq:ufmodel}
\overline{\dot{w}}_c(\underline{x},t)=\bar{\rho}(\underline{x},t)\int_{0}^{1} \frac{\dot{w}_{cL}(\zeta)}{\rho _L (\zeta)}\tilde{p} \left( \zeta;\tilde{c}(\underline{x},t),{\sigma ^2}(\underline{x},t) \right)d\zeta
\end{equation}

\noindent where $\zeta$ is the sample space variable for $c$, $\dot{w}_{cL}$ is the laminar progress variable rate, and $\rho _{L}$ is the laminar density. The progress variable pdf is taken to be a $\beta$-function in accordance to usual practice in presumed pdf methods i.e.,

\begin{equation}
\tilde{p}(\zeta)=\frac{1}{C}{\zeta}^{a-1}(1-\zeta)^{b-1}
\end{equation}

\noindent where $C$ is a normalisation constant and where the parameters $a,b$ are chosen so that the filtered progress variable $\tilde{c}$ and the variance $\sigma ^2$ are recovered. These are given by:
$a=\tilde{c} \left( 1/g-1 \right)$, and $b=(1-\tilde{c})\left( 1/g-1 \right)$, where $g={{\sigma}^2}/{\tilde{c}(1-\tilde{c})}$. 

Note that in practice, the integration as specified by Eq. \ref{eq:ufmodel} may be problematic since the progress variable pdf takes non-finite values in the case the pdf is bimodal. A way around this issue is to expand the integral in Eq. \ref{eq:ufmodel} and use the cumulative distribution function $\tilde{P}(\zeta;\tilde{c},{\sigma ^2})$ instead for calculating the mean of a variable $y$ as, 

\begin{multline}
\overline{y}(\underline{x},t)=
\frac{\bar{\rho}(\underline{x},t)}{\rho _L(1)}y_L(1) \\
-\bar{\rho}(\underline{x},t) \int_0^1 \frac{1}{\rho _L(\zeta)}\left( \rho _L(\zeta) \frac{d y_L(\zeta)}{d \zeta}-y_L(\zeta) \frac{d \rho _L (\zeta)}{d \zeta} \right ) \tilde{P}(\zeta;\tilde{c},{\sigma ^2}) d \zeta
\end{multline}

\noindent where the derivatives of $y$ with respect to $\zeta$ are typically well-defined for quantities of interest, and so is the cumulative distribution function $\tilde{P}$. 

\subsection{Filtered Laminar Flame (FLF) model }\label{sec:flf}

In the FLF approach \cite{moureau_cnf_2010,nambully_cnf_2014}, the filtered rate is calculated from look-up tables constructed using filtered laminar profiles of a canonical flame solution e.g. a 1D unstrained flame. This process essentially corresponds to a filtered laminar flame pdf approach \cite{moureau_cnf_2010,nambully_cnf_2014}, 

%\begin{equation}
%\bar{\dot{w}}_c(\underline{x},t;\Delta)=\bar{\rho}(\underline{x},t;\Delta)\int_{0}^{1}\frac{\dot{w}_{cL}(\zeta)}{\rho _L(\zeta)}\tilde{p}_L(\zeta;\tilde{c};\sigma ^2;\Delta _L)d\zeta
%\end{equation}

\begin{equation}
\bar{\dot{w}}_c(\underline{x},t)=\bar{\rho}(\underline{x},t)\int_{0}^{1}\frac{\dot{w}_{cL}(\zeta)}{\rho _L(\zeta)}\tilde{p}_L \left( \zeta;\tilde{c}(\underline{x},t);\sigma ^2(\underline{x},t) \right) d\zeta
\end{equation}

\noindent where the filtered laminar flame pdf is given by,

%\begin{equation}
%\tilde{p}_L(\zeta;\tilde{c};\sigma ^2;\Delta _L)=\frac{\rho _L(\zeta)}{\rho _L(\tilde{c},\sigma ^2)} \left| \frac{d \zeta}{dx} \right| ^{-1} G\left(x(\tilde{c}, \Delta _L)-x(\zeta,\Delta _L) \right)
%\end{equation}

\begin{equation}
\tilde{p}_L(\zeta;\tilde{c};\sigma ^2)=\frac{\rho _L(\zeta)}{\bar{\rho} ^{\Delta _L} _L} \left| \frac{d \zeta}{dx} \right| ^{-1} G\left({x} ^{\Delta _L} _L-x(\zeta) \right)
\end{equation}

\noindent and the filter function $G$ in this case is in 1D. In the FLF approach, the laminar flame filter size $\Delta _L$ which is used to parameterise the laminar flame pdf, is generally smaller than the actual LES filter size $\Delta$. The laminar flame filter $\Delta _L$, and the laminar flame position $x_L$, are chosen so that the progress variable and its variance as obtained from the LES, match the corresponding 1D-filtered laminar flame values i.e. $\tilde{c}(\underline{x},t;\Delta)=\tilde{c}_L(x_L;\Delta _L)$, and $\sigma ^2(\underline{x},t;\Delta)=\sigma ^2 _L(x_L;\Delta _L)$. This ensures that the filtered-flame pdf recovers the LES mean and variance values. 

In practice, an unstrained laminar flame is filtered for a wide range of filter widths in order to obtain a table of values of $\tilde{c} _L$, and $\sigma ^2 _L$. The LES values $\tilde{c}$ and $\sigma ^2$ are then used in order to determine the corresponding filter width $\Delta _L$ and spatial position $x_L$ in the table which match the filtered laminar flame values. The mean rate is then calculated by filtering the laminar flame profile at $\Delta _L$ and obtaining its value at $x_L$, as discussed in \cite{moureau_cnf_2010,nambully_cnf_2014}. 

\section{Results and Discussion} \label{sec:resdisc}

\subsection{Modeled pdfs comparison}

Irrespective of the flamelet model used, small discrepancies in the modeled variance using deconvolution (or any other variance model) may cause large discrepancies in the modeled pdf and as a result large discrepancies in the filtered progress variable reaction rate. The aim of this section is to investigate this point for the UF and FLF models. 

%PDFS
\begin{figure}[ht!]
%\hspace{-1.0cm}
\subfigure[]{
\includegraphics[scale=0.35, trim=8.0cm 0.0cm 1.5cm 0.0cm]{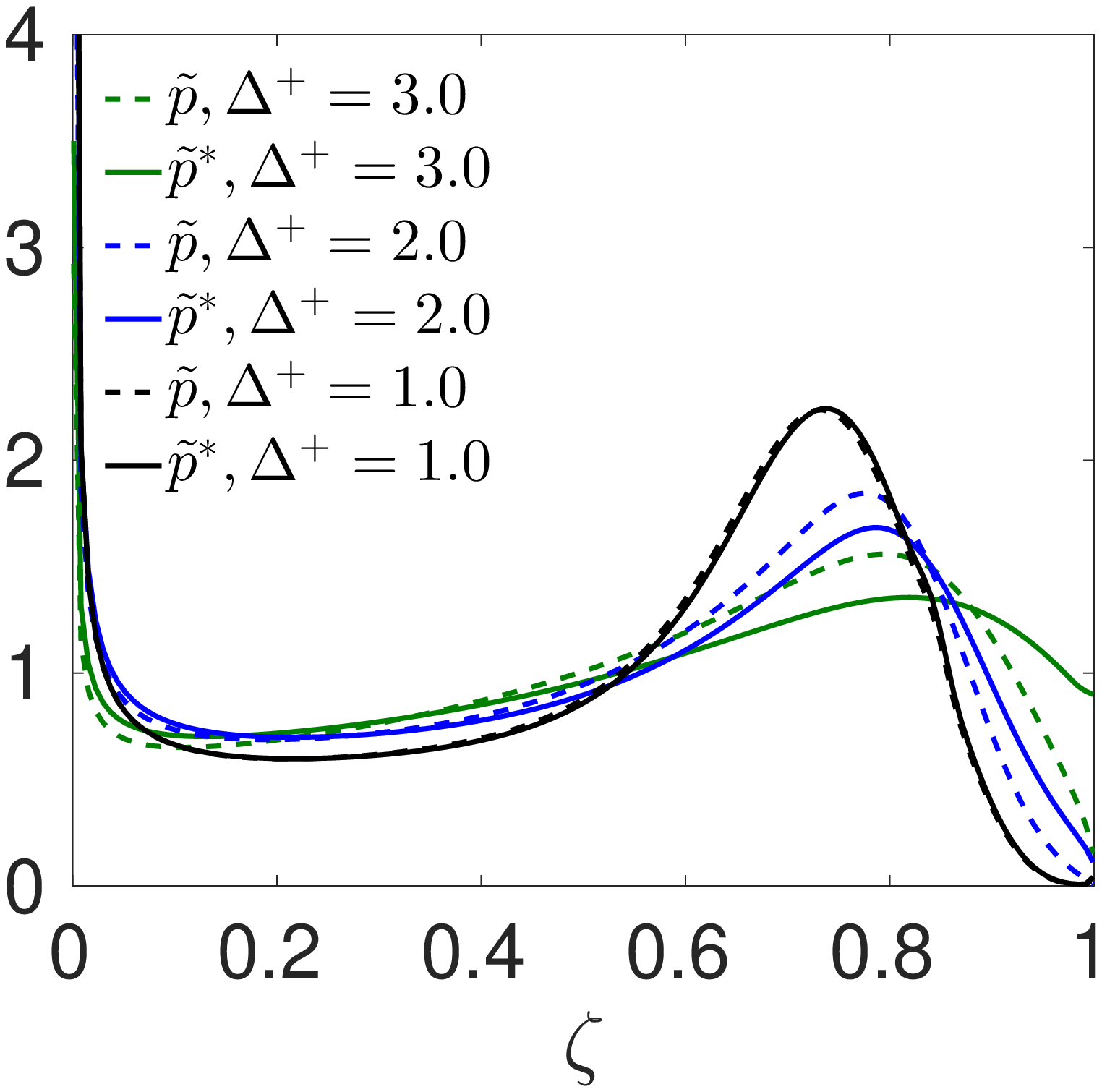}
}
%\hspace{1.0cm}
\subfigure[]{
\includegraphics[scale=0.35, trim=8.0cm 0.0cm 1.5cm 0.0cm]{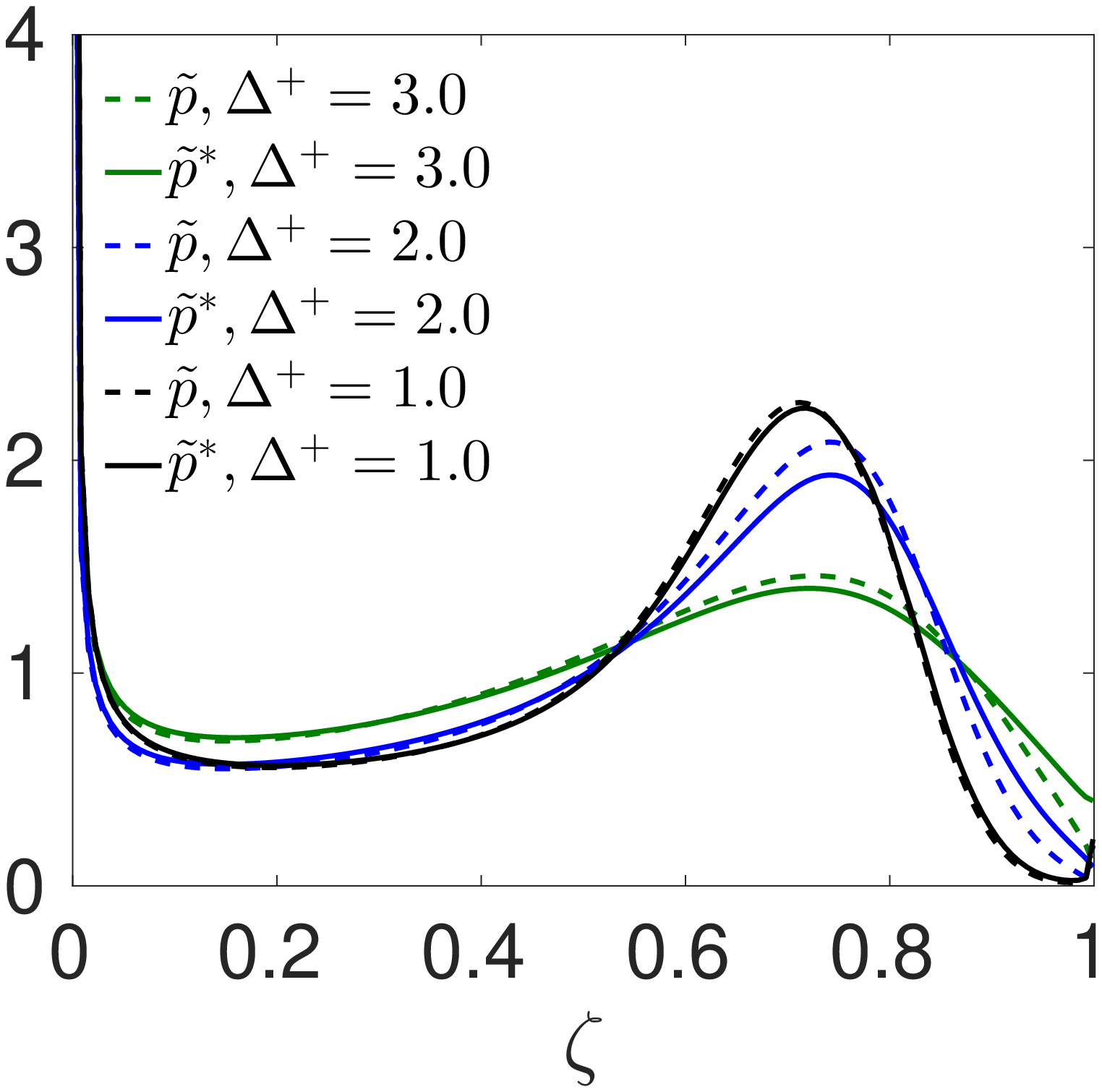}
}
%\hspace{1.0cm}
\subfigure[]{
\includegraphics[scale=0.35, trim=8.0cm 0.0cm 1.5cm 0.0cm]{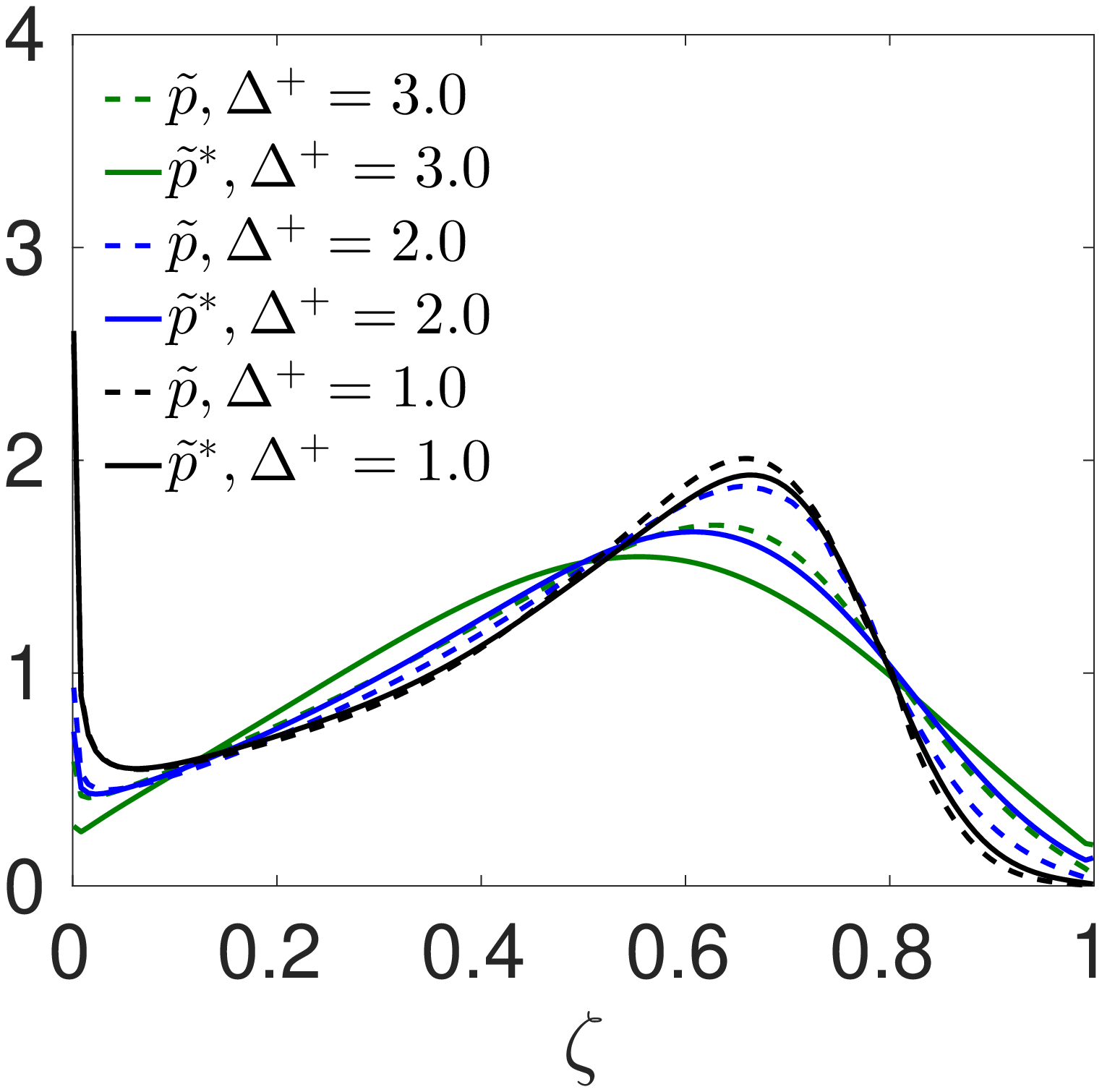}
}
\caption{Progress variables pdfs for the UF model at $<\tilde{c}>$=0.5 for (a) case A, (b) case B, and (c) case C. $p$ is obtained using the actual variance, and $p^*$ is obtained using deconvolution for modeling the variance.}
\label{fig:pdfs}
\end{figure}

%PDFS
\begin{figure}[ht!]
%\hspace{-1.0cm}
\subfigure[]{
\includegraphics[scale=0.35, trim=8.0cm 0.0cm 1.5cm 0.0cm]{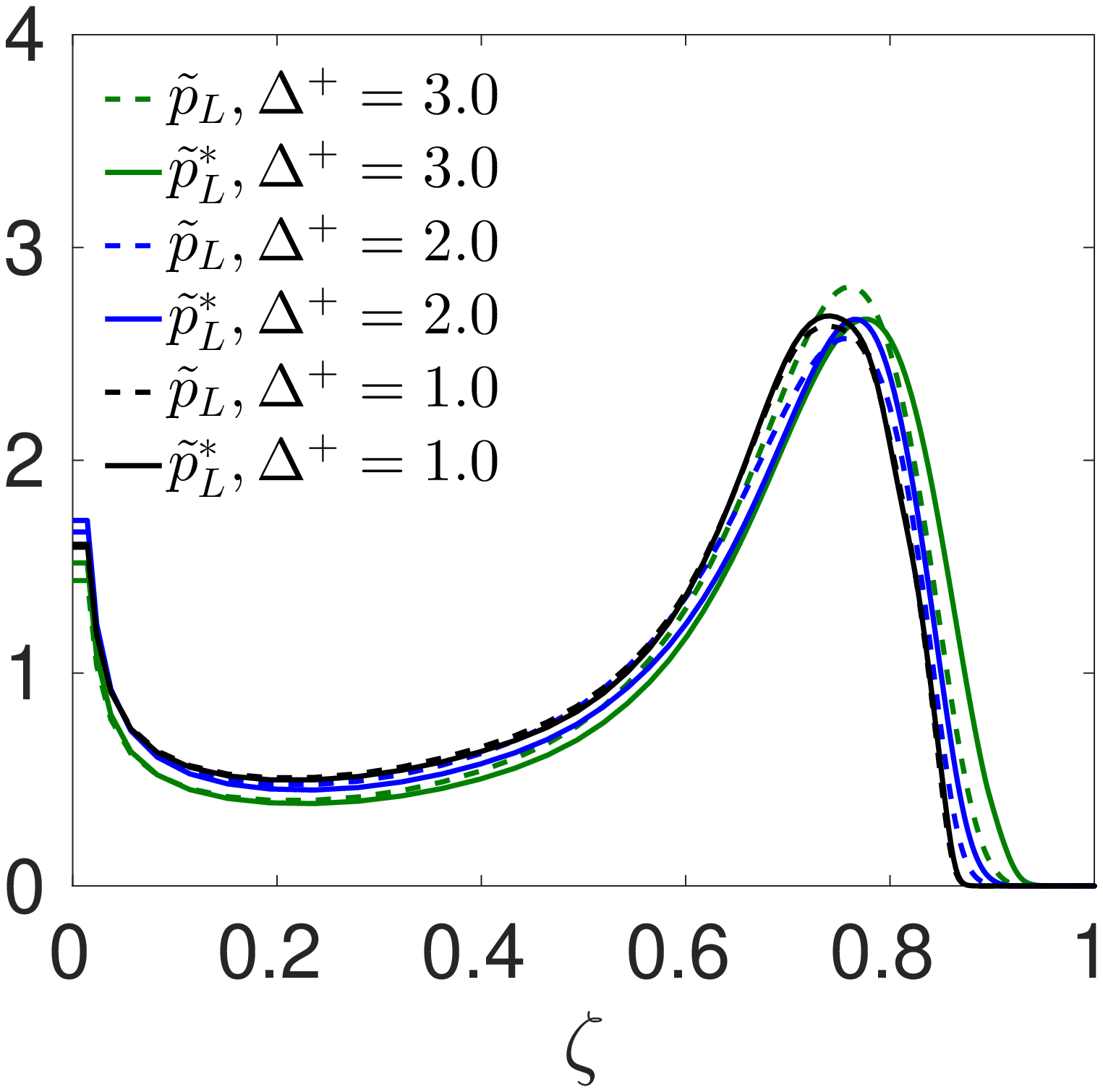}
}
%\hspace{1.0cm}
\subfigure[]{
\includegraphics[scale=0.35, trim=8.0cm 0.0cm 1.5cm 0.0cm]{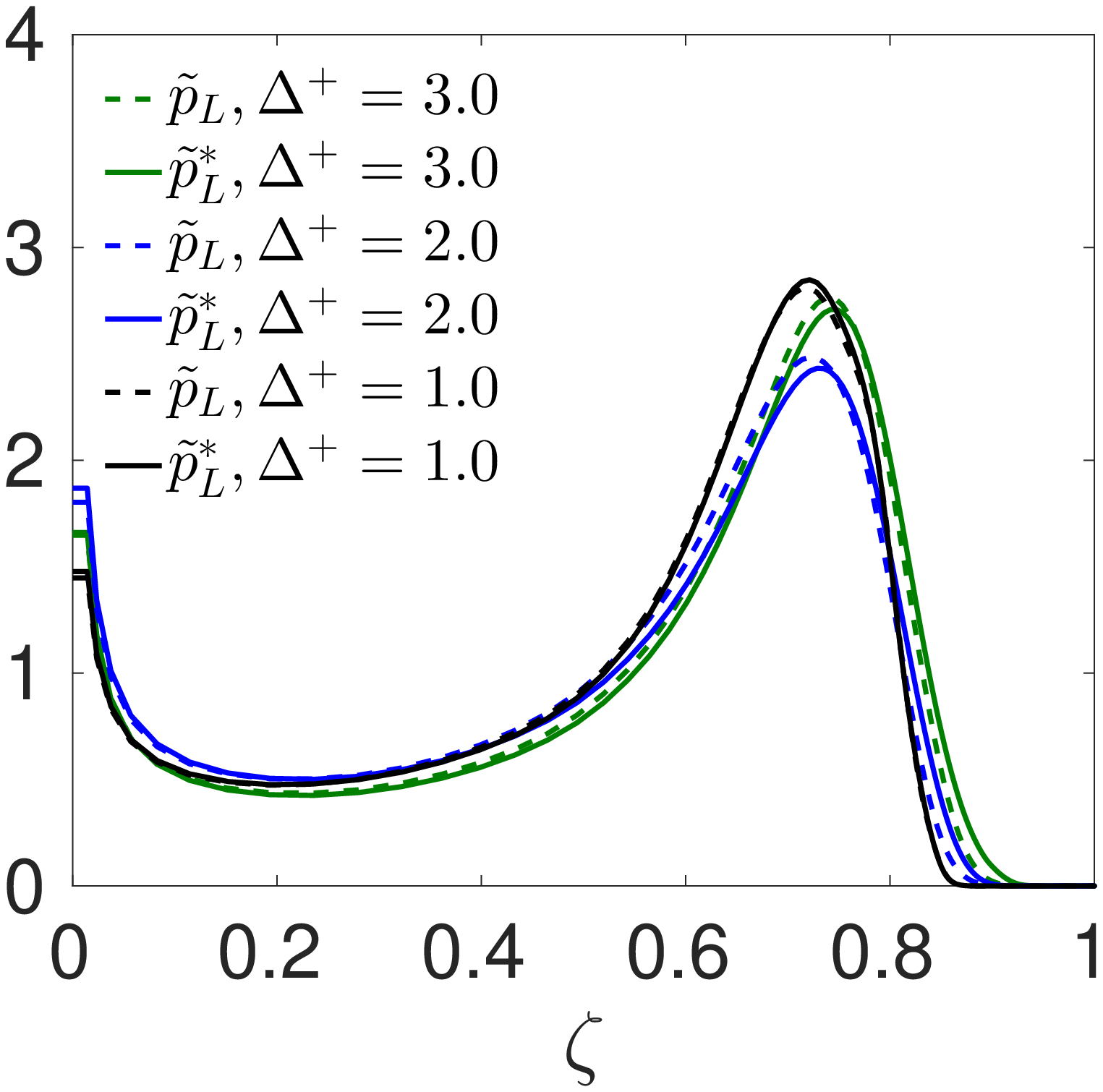}
}
%\hspace{1.0cm}
\subfigure[]{
\includegraphics[scale=0.35, trim=8.0cm 0.0cm 1.5cm 0.0cm]{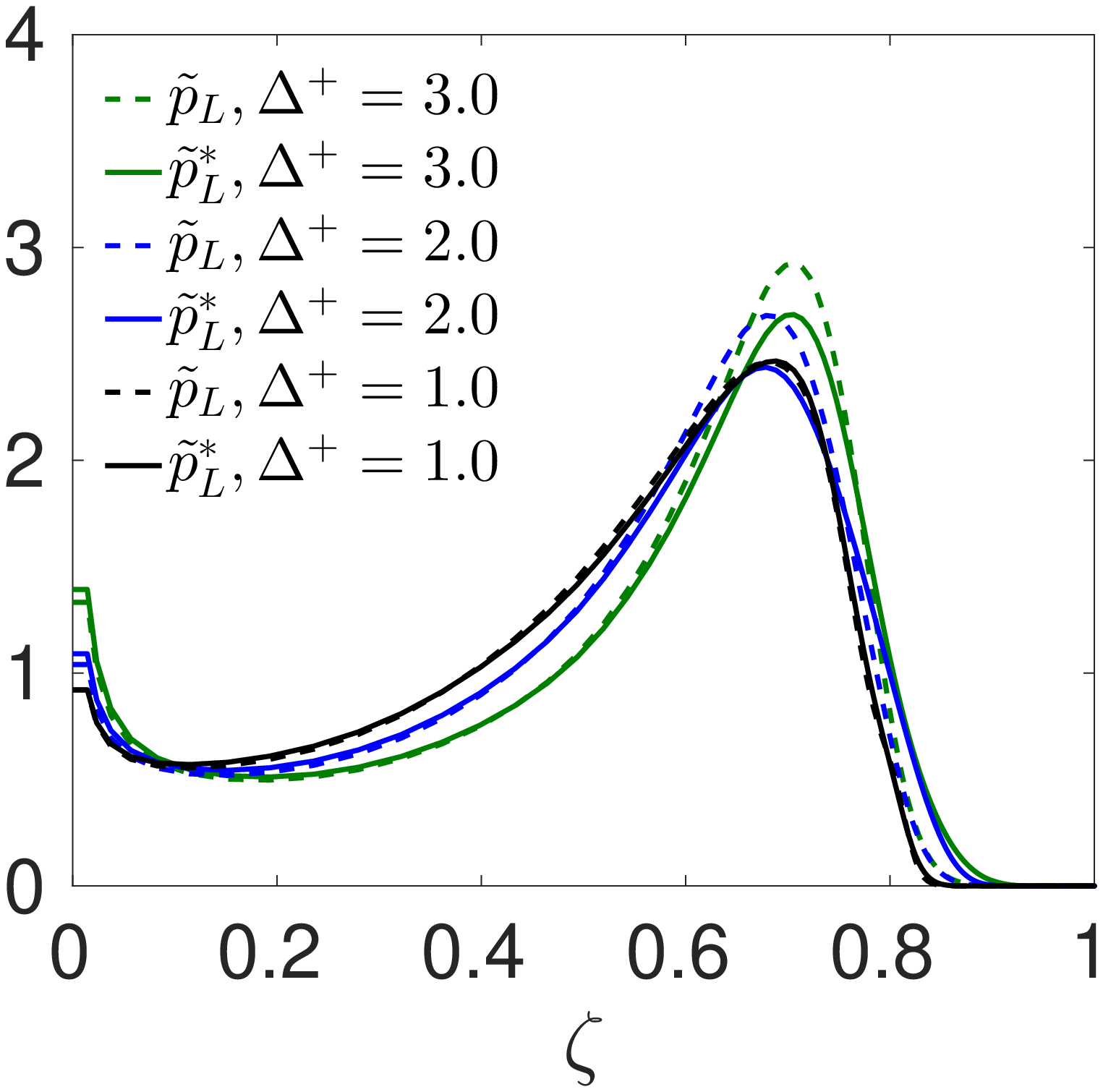}
}
\caption{Progress variables pdfs for the FLF model at $<\tilde{c}>$=0.5 for (a) case A, (b) case B, and (c) case C. $p_L$ is obtained using the actual variance, and $p^* _L$ is obtained using deconvolution for modeling the variance.}
\label{fig:pdfs_FLF}
\end{figure}

In order to do that, the progress variable pdfs, as defined for the UF and FLF models in sections \ref{sec:uf} and \ref{sec:flf} respectively, are calculated at every point in the domain on the LES mesh. The pdfs are parameterised in two different ways: (a) using the actual variance as obtained by filtering the DNS data (and sampling onto the LES mesh) to obtain $p(\zeta)$ and $p_L(\zeta)$, and (b) using the modeled variance as obtained using deconvolution to obtain $p^*(\zeta)$ and ${p^*}_L(\zeta)$. Thus, for every grid-point on the LES mesh, any changes in the modeled pdfs are a result of differences in the variance only, since $\tilde{c}$ is always the same. The pdfs as obtained at every grid-point and for every time-step, are then averaged in the homogeneous directions ($y,z$) and for all time-steps, in order to obtain the averaged $x$-direction pdfs $<p(\zeta;x)>$, $<p^* (\zeta;x)>$ and so on. The Favre-filtered progress variable $\tilde{c}$ is also averaged in the same way to obtain $<\tilde{c}>$ as a function of $x$. This averaging procedure allows us to make a direct comparison between the two modeled pdfs, in order to investigate whether the deconvolution-based modeling for the variance induces any significant bias.  

Figure \ref{fig:pdfs} (a)-(c) shows the averaged pdfs $<p(\zeta)>$ and $<p^* (\zeta)>$ for the UF model for cases A-C respectively. The pdfs at a spatial location corresponding to $<\tilde{c}>$=0.5 are shown, where there is a significant contribution in the burning mode part of the pdf. The agreement between $<p>$ and $<p^ *>$ is best for the lowest filter width, $\Delta ^+$ =1, as expected for all three cases. This result is consistent with the results in \cite{nikolaou_ftc_2018} where the variance estimate is most accurate for the lowest filter widths. For increasing filter widths, the agreement somewhat deteriorates. This is also to be expected since deconvolution accuracy is inversely proportional to filter width size \cite{nikolaou_ftc_2018}. This results in the variance being somewhat over-estimated, hence the generally broader pdfs observed in Fig. \ref{fig:pdfs} (a)-(c) for $\Delta ^+=$2, and, 3. Overall though, the agreement between the two pdfs remains good. 

The same process is repeated for the FLF model. The results for the filtered laminar flame pdfs are shown in Fig. \ref{fig:pdfs_FLF} (a)-(c) for cases A-C respectively. These pdfs also correspond to a spatial position having $<\tilde{c}>$=0.5. 
A similarly good agreement between the modeled pdfs is observed in the case of the FLF model. These results indicate that the deconvolution-based approach for modeling the variance does not introduce a significant bias in the modeled pdfs for either flamelet model used.

%------------------------------------------------------------------

\subsection{Modeled rates comparison}

%Mean rates
\begin{figure}[h!]
\hspace{-2.5cm}
\subfigure[]{
\includegraphics[scale=0.35, trim=8.0cm 0.0cm 8.0cm 0.0cm]{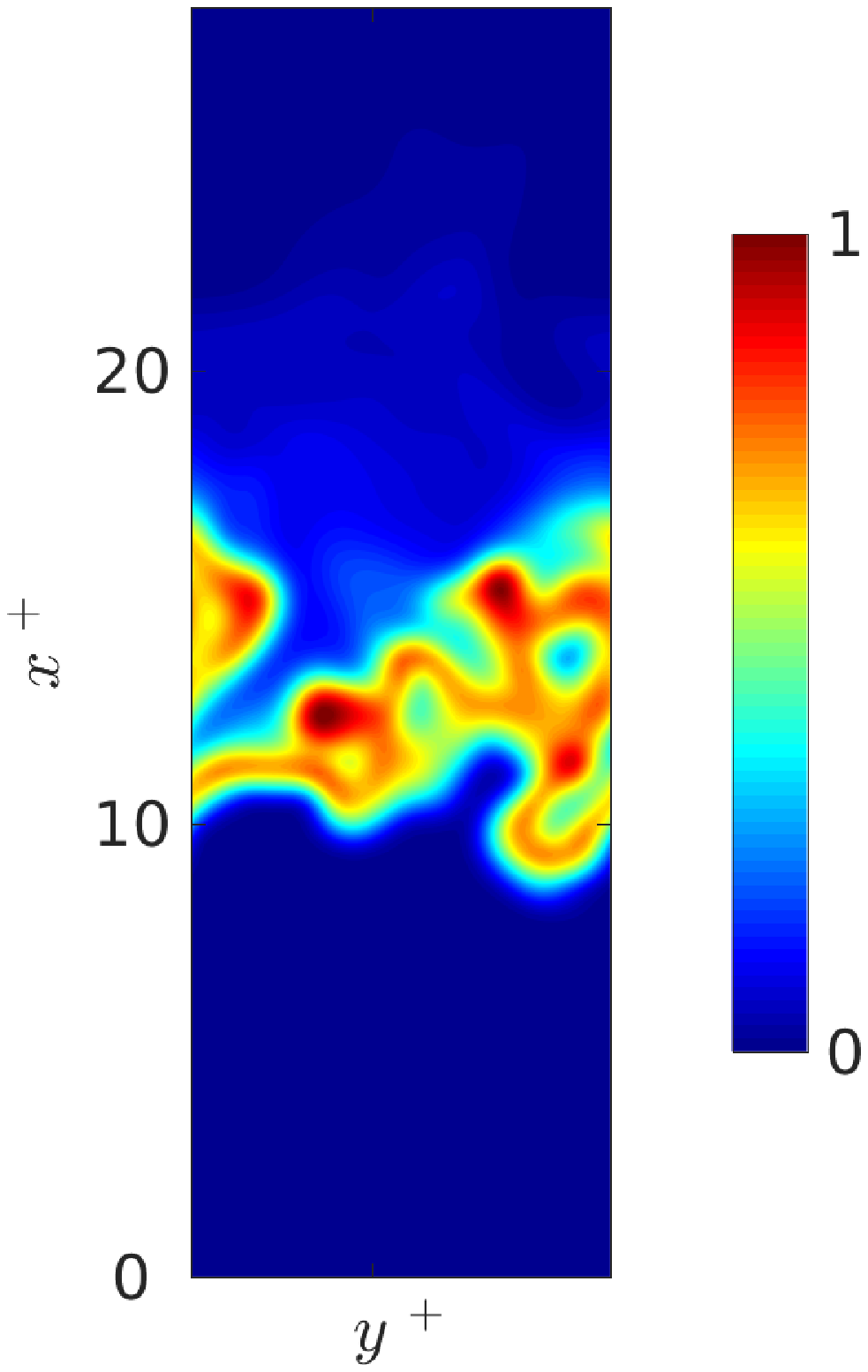} }
%\hfill
\subfigure[]{
\includegraphics[scale=0.35, trim=8.0cm 0.0cm 8.0cm 0.0cm]{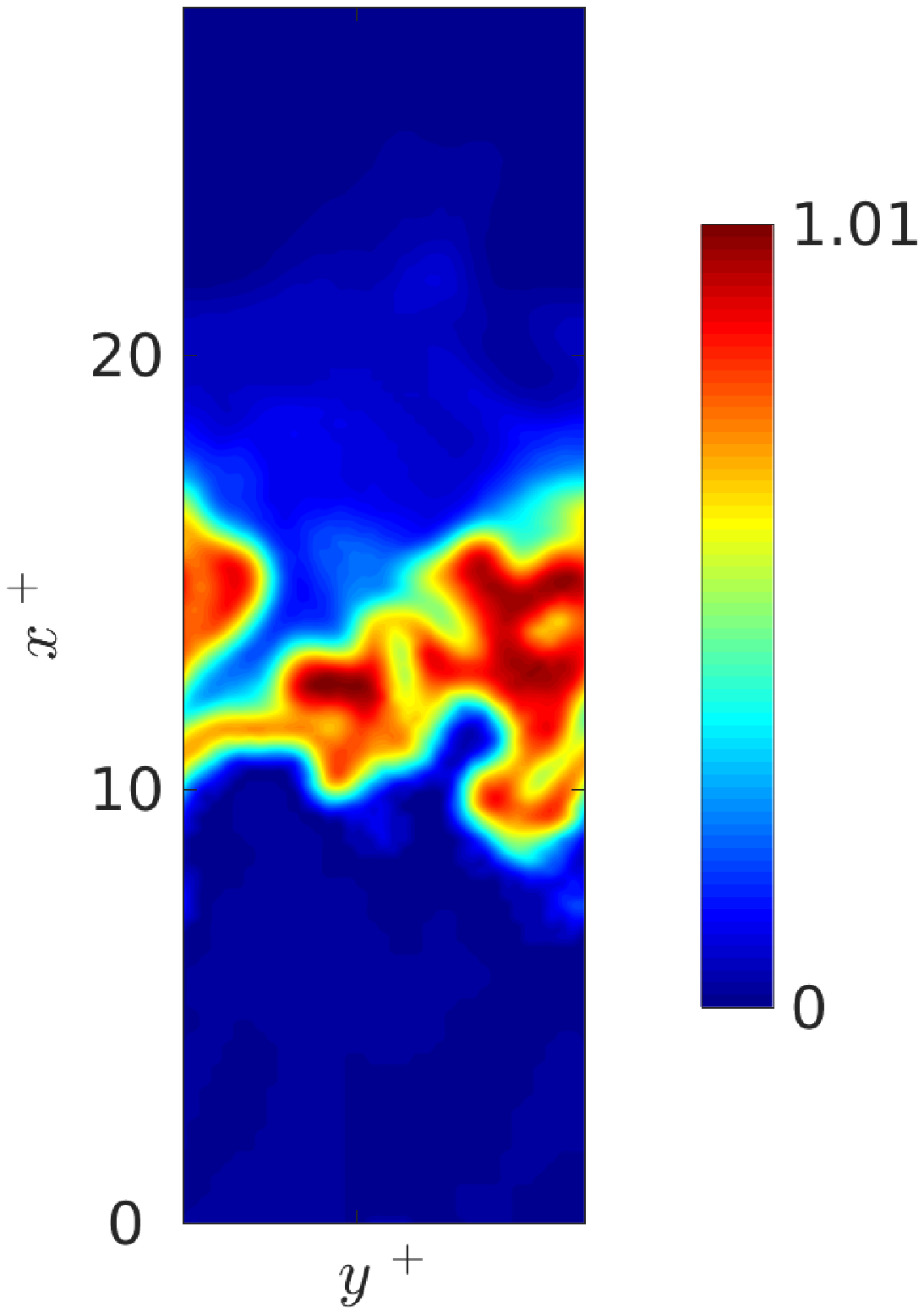} }
%\hfill
\subfigure[]{
\includegraphics[scale=0.35, trim=8.0cm 0.0cm 8.0cm 0.0cm]{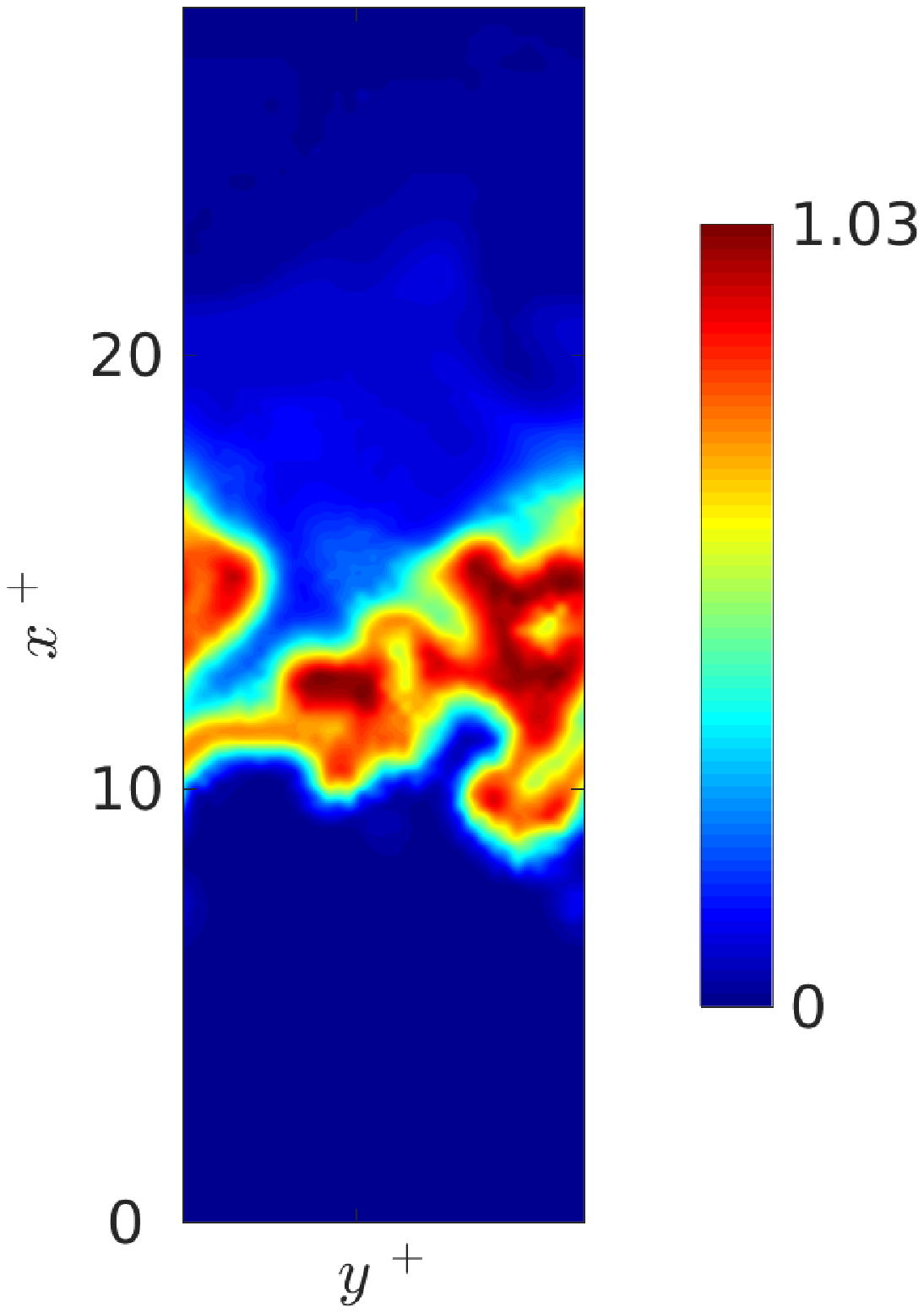} }
\subfigure[]{
\includegraphics[scale=0.35, trim=8.0cm 0.0cm 8.0cm 0.0cm]{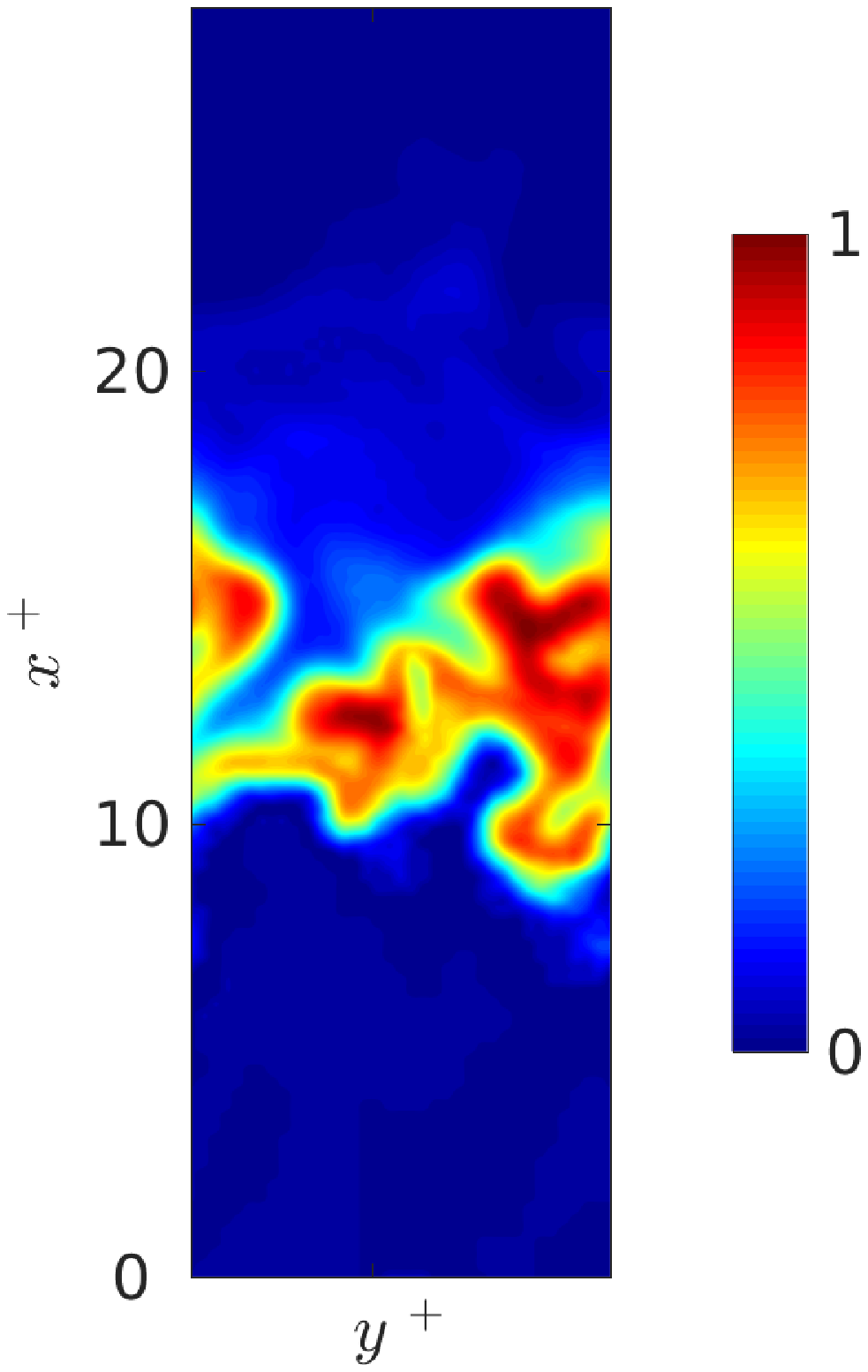} }
\subfigure[]{
\includegraphics[scale=0.35, trim=8.0cm 0.0cm 8.0cm 0.0cm]{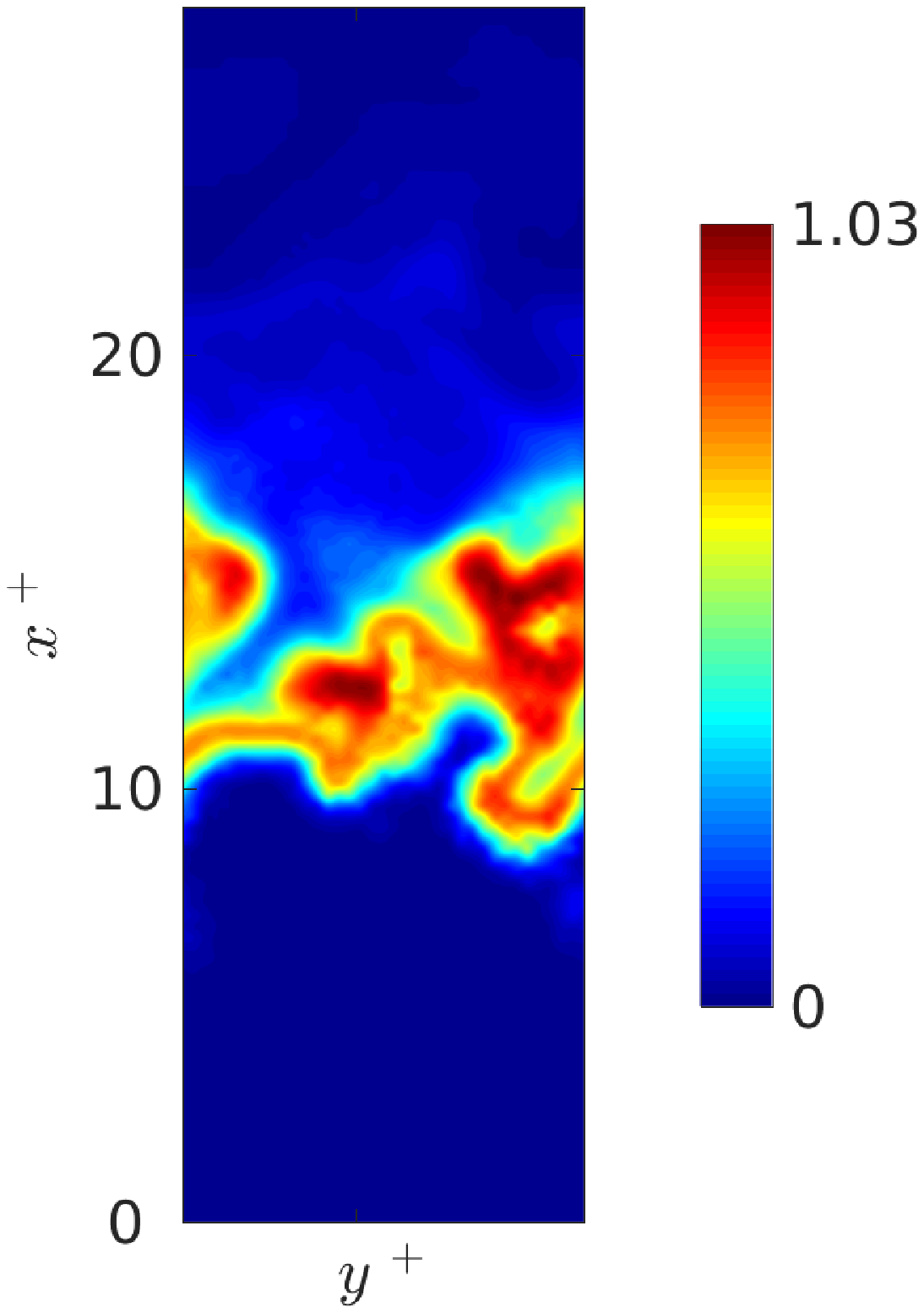} }
\hspace{-2.5cm}
\caption{Comparison of instantaneous progress variable rates for $\Delta ^+=$1.0: (a) $\bar{\dot{w}} _c$, (b) $\bar{\dot{w}} _c:UF$, (c) $\bar{\dot{w}} _c:FLF$, (d) $\bar{\dot{w}} _c:UF-IDEF$, (e) $\bar{\dot{w}} _c:FLF-IDEF$. }
\label{fig:wc_contours}
\end{figure}

The results of the previous section are encouraging, and the aim of this section is to examine whether the hybrid flamelet-deconvolution approach introduces a significant bias in the modeled rates. The pdfs as obtained on the LES mesh are used in order to calculate the progress variable rates for the UF and FLF models (on the LES mesh) as described in sections \ref{sec:uf} and \ref{sec:flf}. The pdfs are calculated using both the actual variance as obtained by filtering the DNS data, and the modeled variance using deconvolution, as explained in the previous section. In the case of the UF model, this leads to filtered rate predictions which are denoted here by $\bar{\dot{w}} _c:UF$, and $\bar{\dot{w}} _c:UF-IDEF$. In the case of the FLF model, the corresponding rate predictions are denoted as $\bar{\dot{w}} _c:FLF$, and $\bar{\dot{w}} _c: FLF-IDEF$. 

Figure \ref{fig:wc_contours} shows a visual comparison between the the actual rate (a) and the four modeled rates (b)-(e) as defined above. This is done for the highest turbulence level, case C, for $\Delta ^+=$1.0. Figure \ref{fig:wc_contours} shows instantaneous contours of the progress variable rate in the direction of mean flame propagation. These contours are equally normalised using the maximum instantaneous value of the DNS-filtered rate in order to elucidate any differences in the modeled rates. Clearly, the flame is highly convoluted due to the intense turbulence level as one may observe from \ref{fig:wc_contours} (a). Nevertheless both the UF (\ref{fig:wc_contours} (b)) and FLF models (\ref{fig:wc_contours} (c)) provide reasonably good estimates of the progress variable rate. In addition, comparing the results in Figs. \ref{fig:wc_contours} (b) and (d) for the UF model it is clear that the deconvolution method provides equally well predictions of the progress variable rate. The same applies for the FLF models by comparing Figs. (c) and (e). Another important point to note is the effect of the filter on the rate: for increasing filter widths the filtered rates generally decrease in comparison with the un-filtered laminar flame result. Simply evaluating the rate using the filtered values would not result in a correct evaluation of the filtered rate since for a non-zero filter width $\bar{\dot{w}}_c(\rho,T,Y_k) \neq {\dot{w}}_c(\bar{\rho},\bar{T},\bar{Y}_k)$. Hence the need for modeling the filtered rate. 

\begin{table}[h!]
\centering
\begin{tabular}{l c c c}
\hline
Case & $\Delta ^+$=1 & $\Delta ^+$=2 & $\Delta ^+$=3  \\ [0.75ex]
\hline
A & 1.2	& 5.7 &	8.8 \\
B & 1.1	& 3.5 &	5.3 \\
C & 1.7 & 6.6 &	10.4 \\
\hline
\hline
\end{tabular}
\caption{Percentage error between modeled rates for the UF model, using IDEF for the pdf variance, and using the actual variance for the pdf, $e(\bar{\dot{w}} _c:UF-IDEF,\bar{\dot{w}} _c:UF)$.}
\label{tbl:error_wc_UF}
\end{table}

\begin{table}[h!]
\centering
\begin{tabular}{l c c c}
\hline
Case & $\Delta ^+$=1 & $\Delta ^+$=2 & $\Delta ^+$=3  \\ [0.75ex]
\hline
A & 6.1	& 9.3 &	11.5 \\
B & 4.9	& 7.6 &	9.2 \\
C & 4.5 & 7.8 &	7.4 \\
\hline
\hline
\end{tabular}
\caption{Percentage error between modeled rates for the FLF model, using IDEF for the pdf variance, and using the actual variance for the pdf, $e(\bar{\dot{w}} _c:FLF-IDEF,\bar{\dot{w}} _c:FLF)$.}
\label{tbl:error_wc_FLF}
\end{table}

\begin{figure}[h]
%\hspace{-1.0cm}
\subfigure[]{
\includegraphics[scale=0.35, trim=8.0cm 0.0cm 0.0cm 0.0cm]{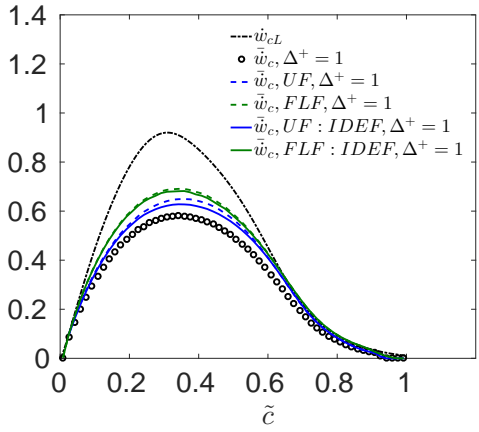}
}
%\hspace{1.0cm}
\subfigure[]{
\includegraphics[scale=0.35, trim=8.0cm 0.0cm 0.0cm 0.0cm]{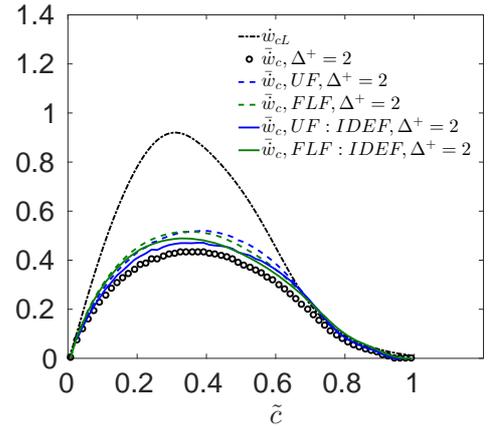}
}
\subfigure[]{
\includegraphics[scale=0.35, trim=8.0cm 0.0cm 0.0cm 0.0cm]{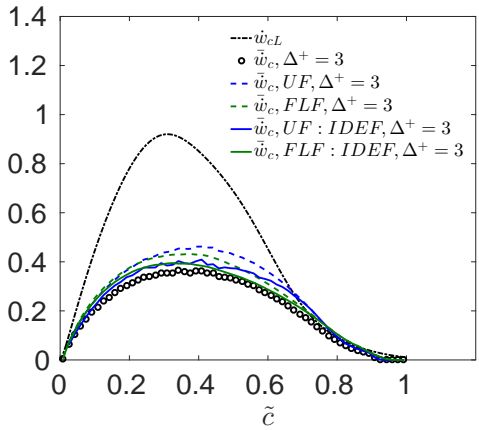}
}
\caption{Conditionally averaged filtered progress variable rate for case A.}
\label{fig:wc_A}
\end{figure}

\begin{figure}[h]
%\hspace{-1.0cm}
\subfigure[]{
\includegraphics[scale=0.35, trim=8.0cm 0.0cm 0.0cm 0.0cm]{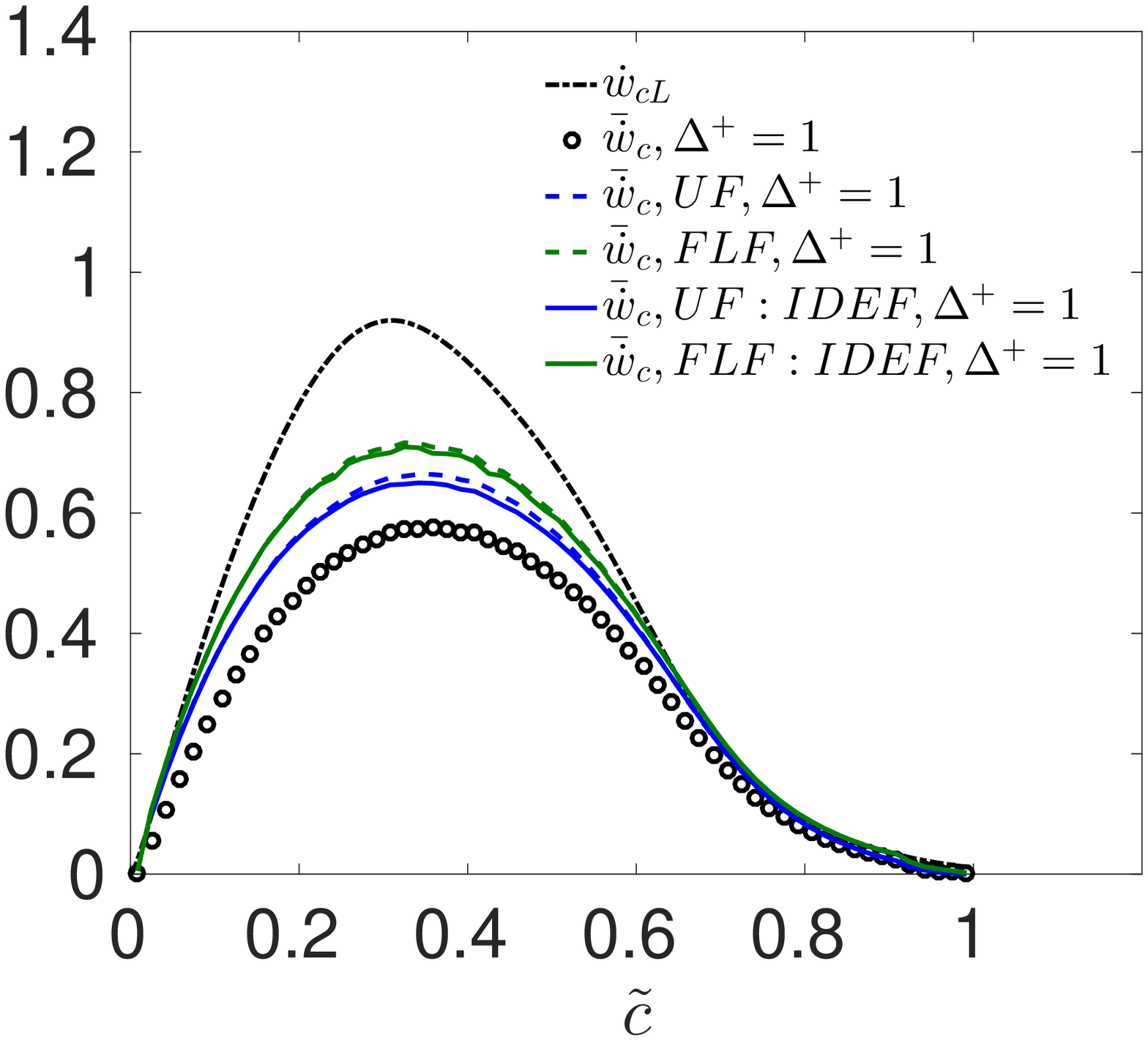}
}
%\hspace{1.0cm}
\subfigure[]{
\includegraphics[scale=0.35, trim=8.0cm 0.0cm 0.0cm 0.0cm]{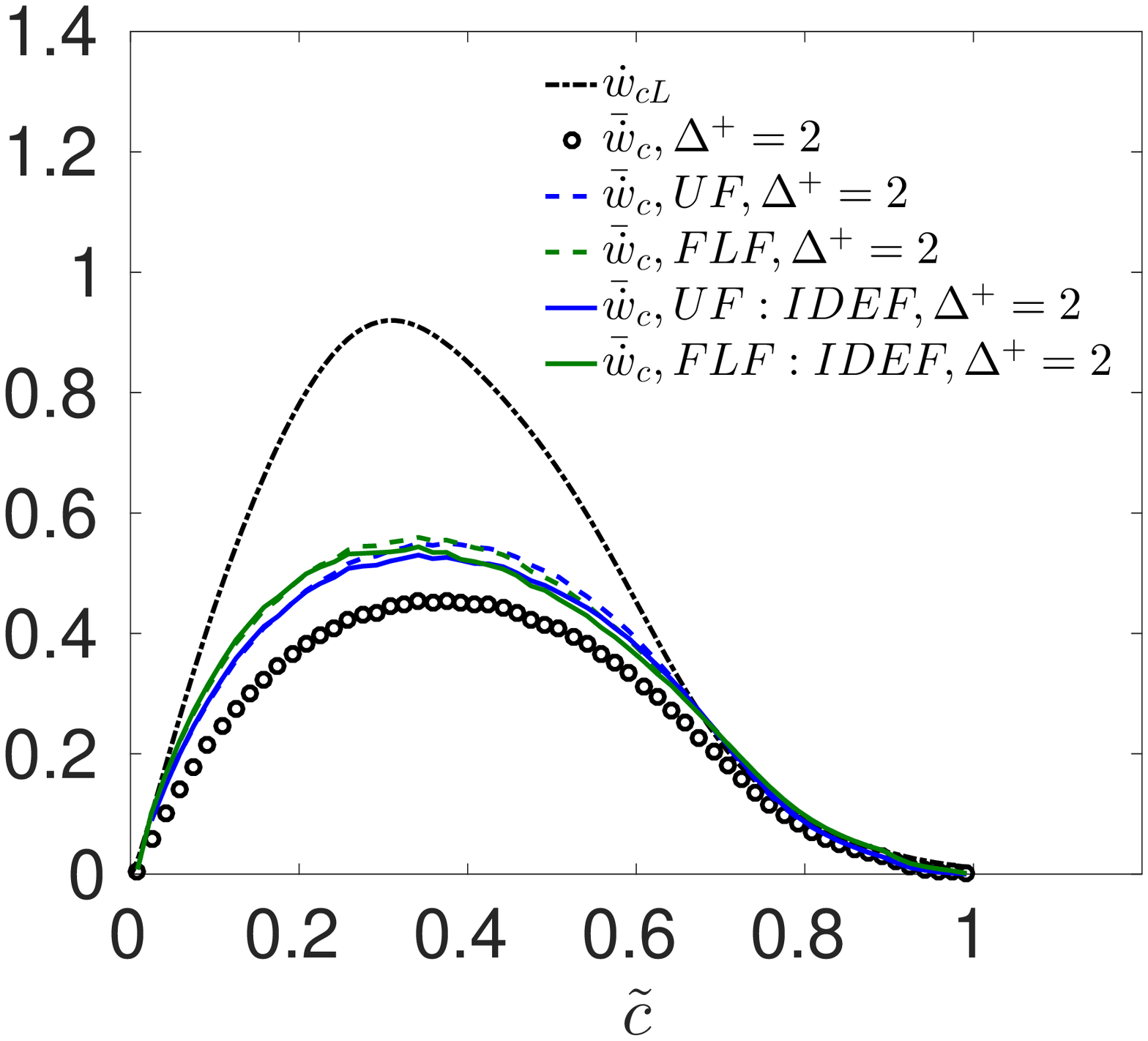}
}
\subfigure[]{
\includegraphics[scale=0.35, trim=8.0cm 0.0cm 0.0cm 0.0cm]{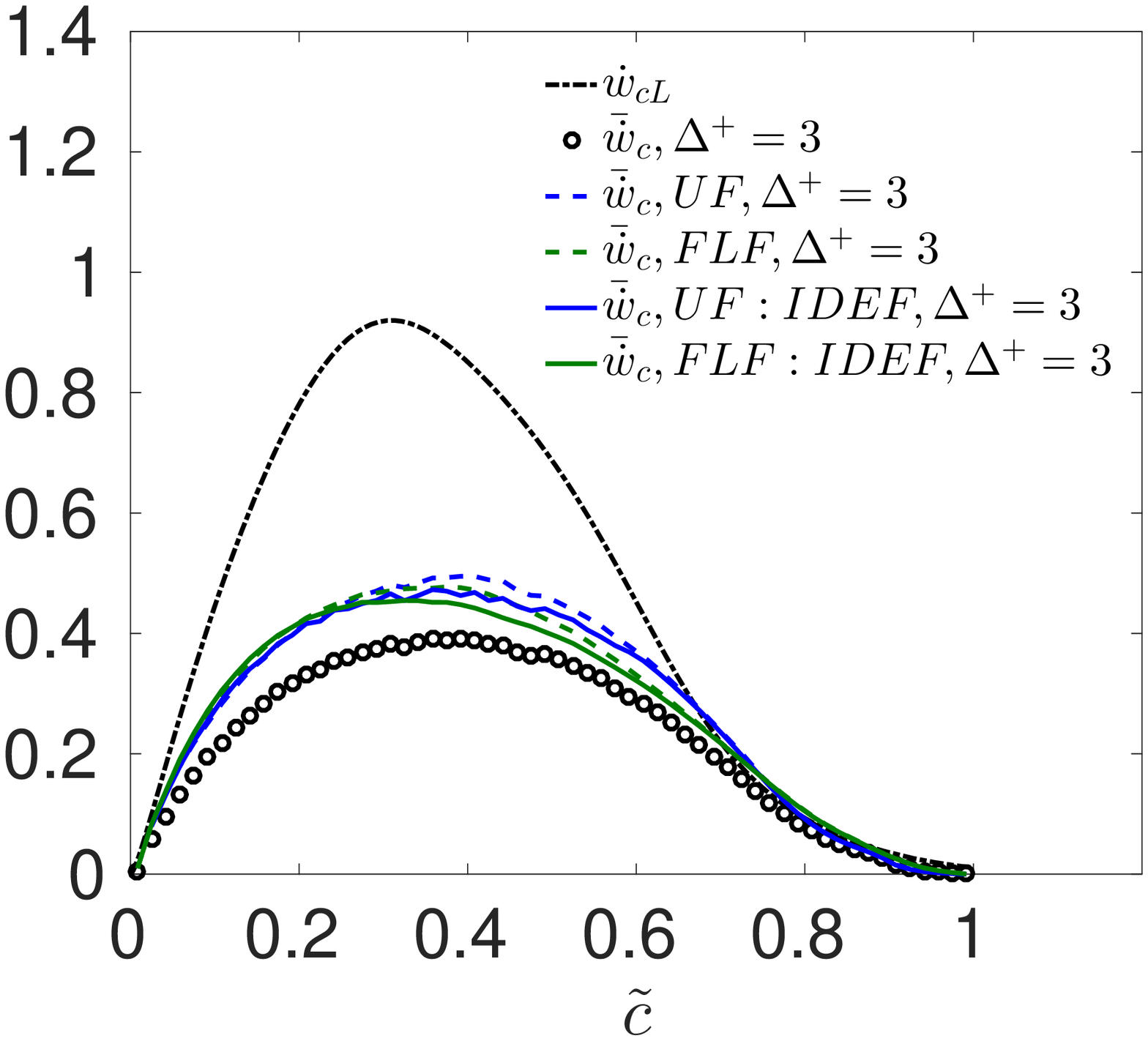}
}
\caption{Conditionally averaged filtered progress variable rate for case B.}
\label{fig:wc_B}
\end{figure}

\begin{figure}[h]
%\hspace{-1.0cm}
\subfigure[]{
\includegraphics[scale=0.35, trim=8.0cm 0.0cm 0.0cm 0.0cm]{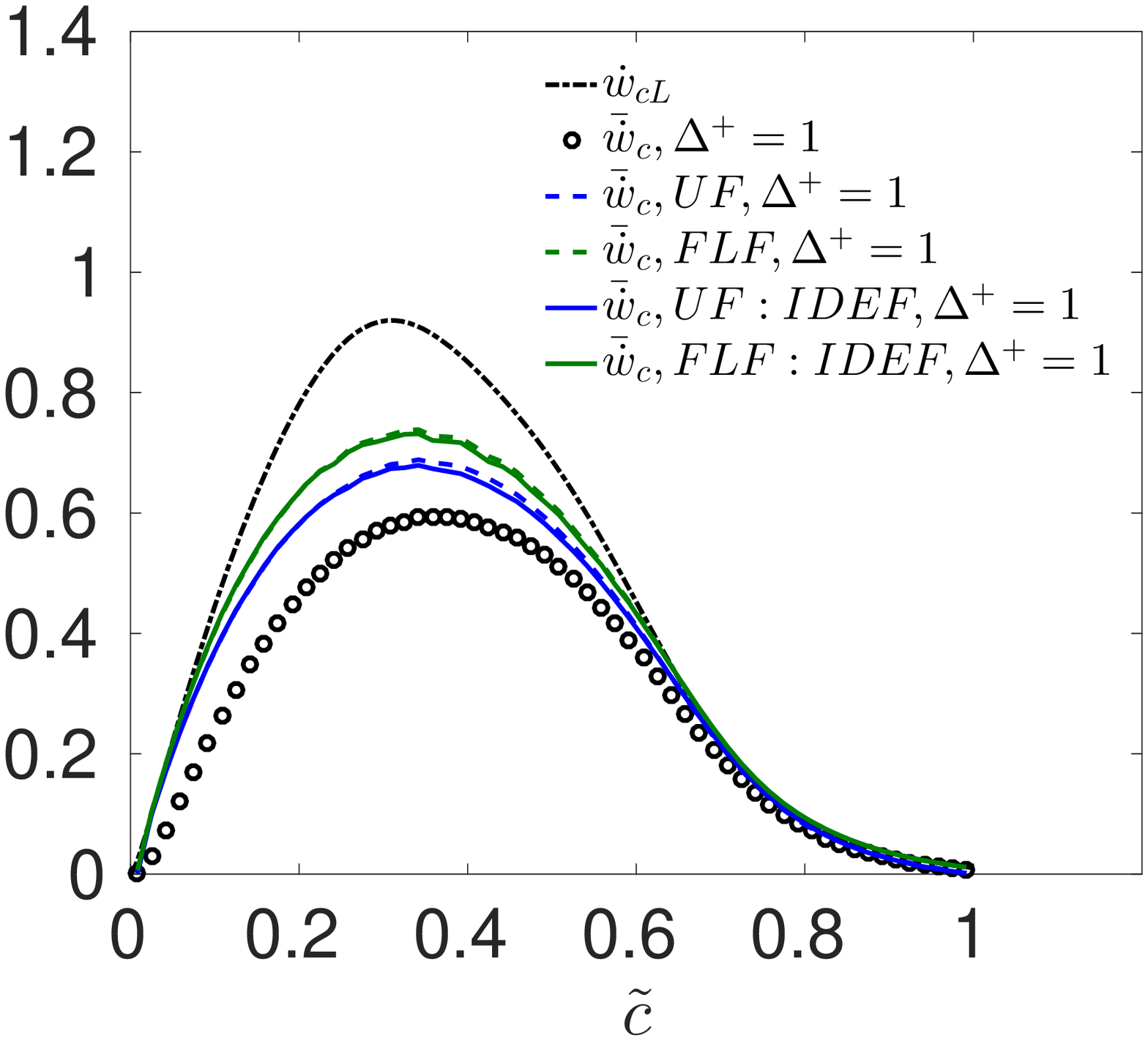}
}
%\hspace{1.0cm}
\subfigure[]{
\includegraphics[scale=0.35, trim=8.0cm 0.0cm 0.0cm 0.0cm]{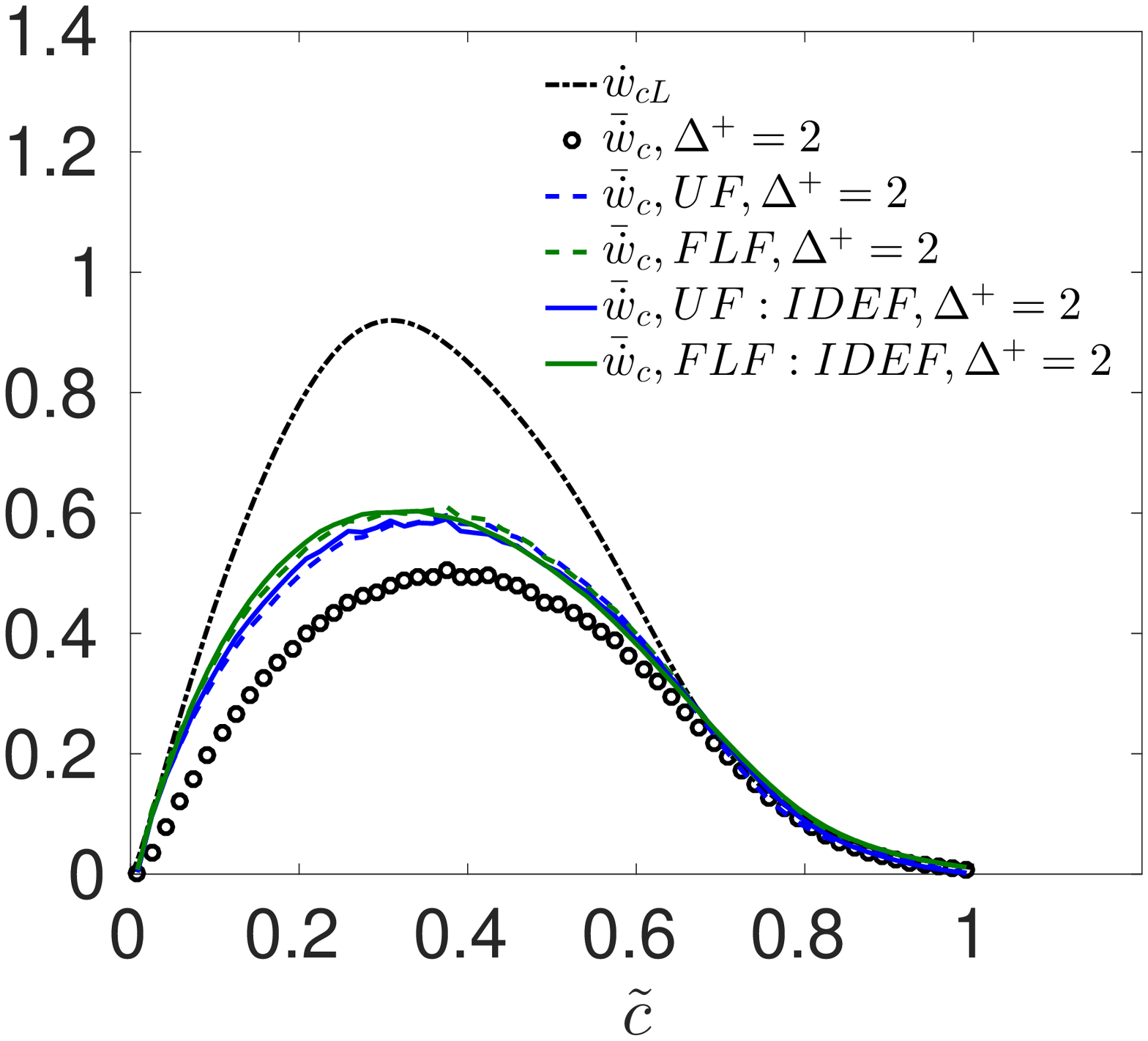}
}
\subfigure[]{
\includegraphics[scale=0.35, trim=8.0cm 0.0cm 0.0cm 0.0cm]{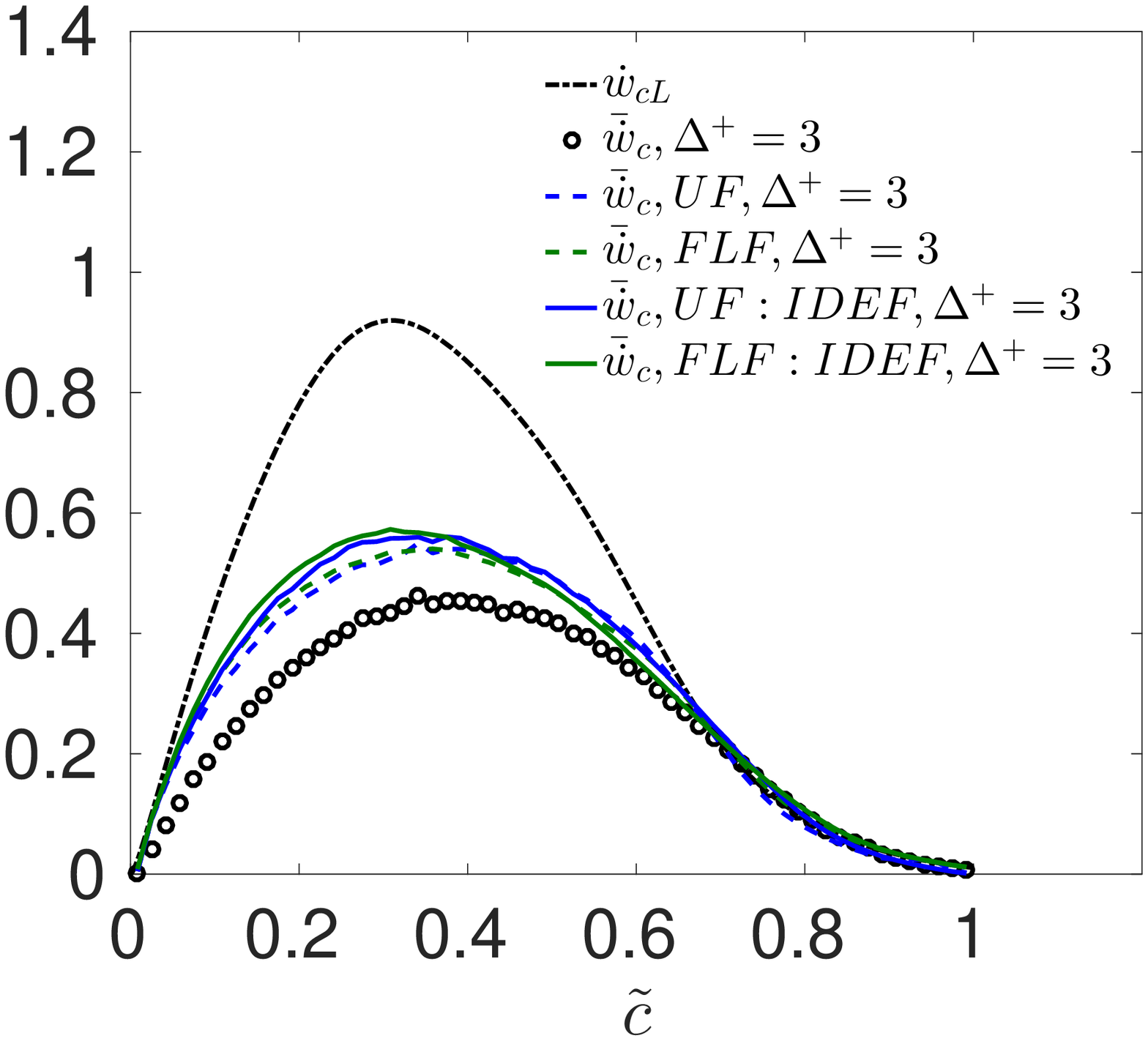}
}
\caption{Conditionally averaged filtered progress variable rate for case C.}
\label{fig:wc_C}
\end{figure}

In order to warrant a more quantitative evaluation, conditional averages based on the Favre-filtered progress variable, $\tilde{c}$, are also calculated for each case, and for each flamelet model. 
The conditional rates for the UF and FLF models for each filter width are shown in Figs. \ref{fig:wc_A}(a)-(c) for case A, Figs. \ref{fig:wc_B}(a)-(c) for case B, and  Figs. \ref{fig:wc_C}(a)-(c) for case C. Also shown in each figure is the laminar flame result, and the actual conditional rate as obtained by filtering the data on the DNS mesh (and sampling onto the LES mesh). The assessment is also complemented by using a suitable local error measure between the modeled rates. This error is defined as $e(x,y)$=100$|x-y| /|y|,|y|>=y_{thr}$ for two variables $x$ and $y$. In the case of the UF model, the errors $e(\bar{\dot{w}} _c:UF,\bar{\dot{w}} _c: UF-IDEF)$ are calculated in order to elucidate any deviations from the UF model predictions. In the case of the FLF model, a similar error is calculated $e(\bar{\dot{w}} _c:FLF,\bar{\dot{w}} _c:FLF-IDEF)$. For each error measure, the threshold is taken to be at 10\% of the maximum instantaneous value of the progress variable rate, so that regions of significant heat release are considered while regions with low heat release do not affect the statistics. These errors are shown in Tables \ref{tbl:error_wc_UF} and \ref{tbl:error_wc_FLF} for the UF and FLF models respectively.

The results in Figs. \ref{fig:wc_A}, \ref{fig:wc_B} and \ref{fig:wc_C} show a reasonably good agreement between the modeled rates both for the UF and FLF approaches. This is also supported by the results in Tables \ref{tbl:error_wc_UF} and \ref{tbl:error_wc_FLF}. For both models the bias in using deconvolution for modeling the variance is lowest for the smallest filter width which is to be expected. The bias generally increases for increasing filter width for each case, but remains within reasonable bounds for the largest filter width. It is interesting to note that the largest error for the UF model occurs for case C at $\Delta ^+=$3.0, while in the case of the FLF model this occurs for case A instead, at $\Delta ^+=$3.0. There does not appear to be a correlation in the error with increasing turbulence level for either the UF or FLF models-further parametric DNS studies are required for determining such behaviour in order to cover a wider space of possible progress variable variance values. What these results clearly indicate however is that IDEF produces quantitatively good estimates of the progress variable rate, with no significant bias in the models' predictions.

\clearpage
\newpage
\section{Conclusions}

A direct numerical simulation database of a freely-propagating turbulent premixed flame in a canonical inflow-outflow configuration, is used in order to investigate whether a deconvolution-based flamelet approach for modeling the filtered progress variable reaction rate in the context of LES, introduces any significant bias in the predictions of the flamelet models used.

The assessment of the method is conducted a priori, on a simulated LES mesh which is constructed using the DNS-filtered data. On the LES mesh, the filtered fields $\bar{\rho}$ and  $\overline{\rho c}$ are deconvoluted, and the progress variable variance is obtained by explicitly filtering the deconvoluted fields on the same mesh. The variance as obtained, is used to parameterise the progress variable pdfs in two popular flamelet methods namely the unstrained flamelet model using a $\beta$-pdf, and the filtered laminar flame model. 

The results of this study show that the deconvolution-based approach for modeling the variance does not introduce any significant bias in the shape of the modeled progress variable pdfs or on the values of the modeled reaction rates for both flamelet models tested. As a result, deconvolution can be used in conjuction with such flamelet methods for modeling the progress variable reaction rate, thus rendering a more parameter-free flamelet-based modeling strategy.  

\vspace{0.5cm}

\textbf{Conflict of interests:} The Authors declare that they have no conflict of interest.

\end{document}